\documentclass[pra,aps,twocolumn,showpacs]{revtex4-1} 
\usepackage{amsmath,amssymb,amsfonts,amsthm}
\usepackage{mathrsfs}
\usepackage{extarrows}
\usepackage{graphicx}
\usepackage{subfigure}
\usepackage{bm}
\usepackage{float}
\usepackage{color}
\usepackage{sidecap}
\allowdisplaybreaks
\usepackage{soul}
\setstcolor{red}
\usepackage[normalem]{ulem}
\usepackage{hyperref}
\hypersetup{colorlinks=true,breaklinks,urlcolor=blue,linkcolor=blue,citecolor=blue}

\begin{document}

\title{Topological phases and Anderson localization in off-diagonal mosaic lattices}
\author{Qi-Bo Zeng}
\email{zengqibo@cnu.edu.cn}
\affiliation{Department of Physics, Capital Normal University, Beijing 100048, China}
\author{Rong L\"u}
\affiliation{State Key Laboratory of Low Dimensional Quantum Physics, Department of Physics, Tsinghua University, Beijing 100084, China}
\affiliation{Frontier Science Center for Quantum Information, Beijing 100084, China}
\begin{abstract}
We introduce a one-dimensional lattice model whose hopping amplitudes are modulated for equally spaced sites. Such  mosaic lattice exhibits many interesting topological and localization phenomena that do not exist in the regular off-diagonal lattices. When the mosaic modulation is commensurate with the underlying lattice, topologically nontrivial phases with zero- and nonzero-energy edge modes are observed as we tune the modulation, where the nontrivial regimes are characterized by quantized Berry phases. If the mosaic lattice becomes incommensurate, Anderson localization will be induced purely by the quasiperiodic off-diagonal modulations. The localized eigenstate is found to be centered on two neighboring sites connected by the quasiperiodic hopping terms. Furthermore, both the commensurate and incommensurate off-diagonal mosaic lattices can host Chern insulators in their two-dimensional generalizations. Our work provides a platform for exploring topological phases and Anderson localization in low-dimensional systems. 
\end{abstract}
\maketitle
\date{today}

\section{Introduction}
During the past two decades, topological phases have drawn considerable attention from almost every research field of physics~\cite{Hasan2010RMP,Qi2011RMP,Ando2015ARCMP,Elliott2015RMP}. To get a better understanding and description of the topological phases, various tight-binding models, or lattice models, have been constructed; for instance, the Haldane model for quantum Hall effect~\cite{Haldane1988PRL}, the Kane-Mele model for quantum spin Hall effect~\cite{Kane2005PRL}, the one-dimensional (1D) Kitaev model for topological superconductors~\cite{Kitaev2001}, and so on. Among all these models, modulations present in the on-site potentials and hopping amplitudes, which are called diagonal and off-diagonal modulations accordingly, play a decisive role in determining the systems' properties. Apart from the spatial modulations varying in configuration space, there are also temporal modulations which are applied to systems periodically, e.g. Floquet systems~\cite{Eckardt2017RMP,Rudner2020NRP}. With the presence of appropriate symmetries, many exotic topological phases can be realized by tuning the modulations, such as topological insulators~\cite{Bernevig2006Science}, Weyl/Dirac semimetals~\cite{Hasan2017ARCMP,Yan2017ARCMP,Armitage2018RMP}, nodal-line semimetals~\cite{Fang2016CPB}, and more recently, higher-order topological insulators~\cite{Fritz2012PRL,ZhangFan2013PRL,Taylor2017Science} and non-Hermitian topological phases~\cite{Bergholtz2021RMP}.

In addition to topological phases, different types of modulation can also influence the systems' localization properties significantly~\cite{Anderson1958PR,Lee1985RMP,Abrahams2010}. One paradigmatic example is the 1D Aubry-Andr\'e-Harper (AAH) model with on-site quasiperiodic modulations, where Anderson localization phase transition emerges as the modulation becomes stronger than a critical value~\cite{Harper1955,Aubry1980}. Various generalizations of the AAH model have been introduced to study the Anderson localization in 1D quasicrystals\cite{Siggia1983PRL,Kohmoto1983PRL,DasSarma1988PRL,DasSarma1990PRB,DasSarma2009PRA,DasSarma2010PRL,
Ganeshan2015PRL,Liu2018PRB,Deng2019PRL}. In addition, the AAH model also exhibits interesting topological properties~\cite{Kraus2012PRL,Ganeshan2013PRL,Liu2015PRB,Zeng2020PRB}. Recently, the 1D mosaic lattice is introduced, where quasiperiodic modulations are added to the sites with equally spaced distance and Anderson localization phenomena with multiple mobility edges are observed~\cite{Wang2020PRL}. By further including such diagonal mosaic modulations to the 1D topological superconductors mode, robust nontrivial phases are reported~\cite{Zeng2020arxiv}. However, the effect of off-diagonal mosaic modulations on the topological phases and Anderson localization has been ignored and remains illusive so far.

In this paper, we propose a 1D lattice model with modulations purely added to the hopping amplitudes for equally spaced sites, which is called off-diagonal mosaic lattice and exhibits interesting properties in topological phases and Anderson localization. When the lattice is commensurate, zero- as well as nonzero energy topological edge modes are observed at the two ends of the lattice by tuning the amplitude of the modulation. The topologically nontrivial phases are characterized by quantized Berry phases. On the other hand, if the mosaic modulation becomes incommensurate with the underlying lattice, the Anderson localization phenomenon emerges. The localization is induced purely by the quasiperiodic off-diagonal modulation and the eigenstate is found to be centered on two neighboring sites with (un)equal distribution weight, which are connected by the mosaic modulated hopping terms and acts as impurities in the lattice. Furthermore, we find that by generalizing the model into a two-dimensional (2D) system, Chern insulators can be obtained in both the commensurate and incommensurate cases. Our model provides a rich playground for studying topological phases and Anderson localization by varying the off-diagonal modulations. 

The rest of the paper is organized as follows. In Sec.~\ref{sect2} we introduce the model Hamiltonian of the 1D off-diagonal mosaic lattices. Then we explore the topological phases in the commensurate lattices in Sec.~\ref{sect3}. The quasiperiodic case with incommensurate modulations is checked in Sec.~\ref{sect4}. The last section (Sec.~\ref{sect5}) is dedicated to a brief summary.

\begin{figure}[t]
\includegraphics[width=2.5in]{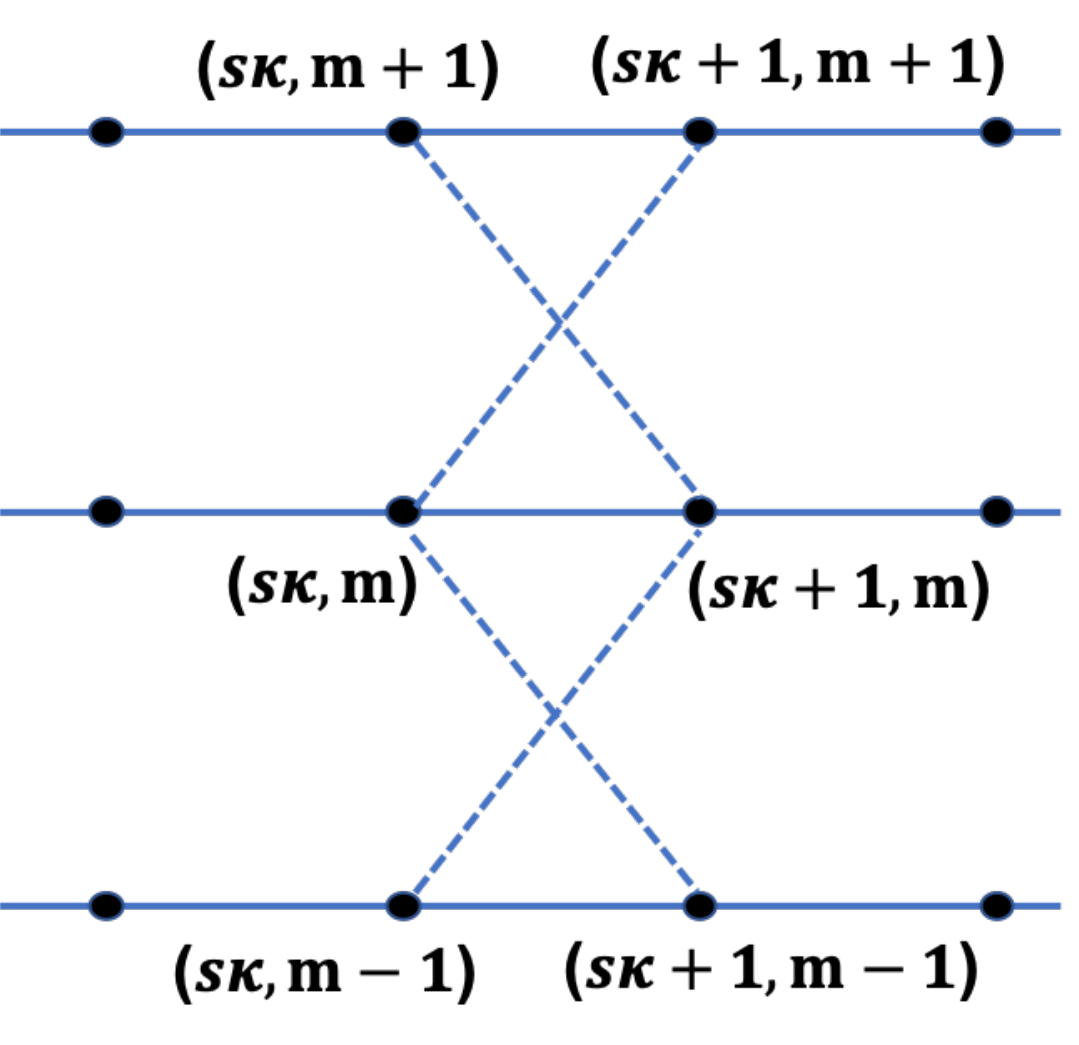}
\caption{(Color online). Schematic illustration of the 2D generalization of 1D mosaic off-diagonal lattice. The black dots represent the lattice sites. The solid line indicates the constant hopping $t$, and the dashed line represents the hopping amplitude due to the mosaic modulation in the 1D model.}
\label{fig1}
\end{figure}

\section{Model Hamiltonian}\label{sect2}
The 1D off-diagonal mosaic lattice model is described by the following Hamiltonian
\begin{equation}\label{H}
H=\sum_{j=1}^{L-1} t_j c_{j+1}^\dagger c_j + H.c.
\end{equation}
This 1D chain has $L$ sites and $c_j^\dagger$ ($c_j$) is the creation (annihilation) operator of spinless fermion at site $j$. The hopping amplitude between the nearest neighboring sites $t_j$ is modulated as
\begin{equation}\label{tj}
t_j =\left\lbrace 
\begin{aligned}
& t + \lambda \cos(2 \pi \alpha j + \phi),\qquad j = s \kappa, \\
& t,\qquad \text{otherwise}.
\end{aligned}
\right. 
\end{equation}
Throughout the paper, we will set $t=1$ as the energy unit. $\lambda$ denotes the strength of the modulation and $s=1,2,3,\cdots$. The modulation in the hopping term occurs with interval $\kappa$. If $\kappa=1$, the model reduces to the regular lattice with modulations applied to every hopping term, which is called off-diagonal AAH model. With nonzero $\phi$, the model in Eq. (\ref{H}) can also be generated from a 2D Hamiltonian, which is obtained by taking $\phi$ as the momentum along another direction, e.g. the $y$ direction. In real space, we have 
\begin{align*}
&H_{2D}=\sum_{j,m} t (c_{j+1,m}^\dagger c_{j,m} + h.c.) \\
&+ \sum_{j=\kappa s,m} \frac{\lambda}{2} [ e^{i2\pi \alpha j} (c_{j+1,m}^\dagger c_{j,m+1} + c_{j,m}^\dagger c_{j+1,m+1}) + h.c. ].
\end{align*}
As illustrated in Fig.~\ref{fig1}, there is no hopping in the vertical direction. The mosaic modulation is represented by the dashed line in the figure and only happens when $j=s\kappa$. In the following, we will investigate the influence of mosaic modulations on the topological phase and Anderson localization.

\section{Topological phases in commensurate lattices}\label{sect3}
\begin{figure}[t]
\includegraphics[width=3.4in]{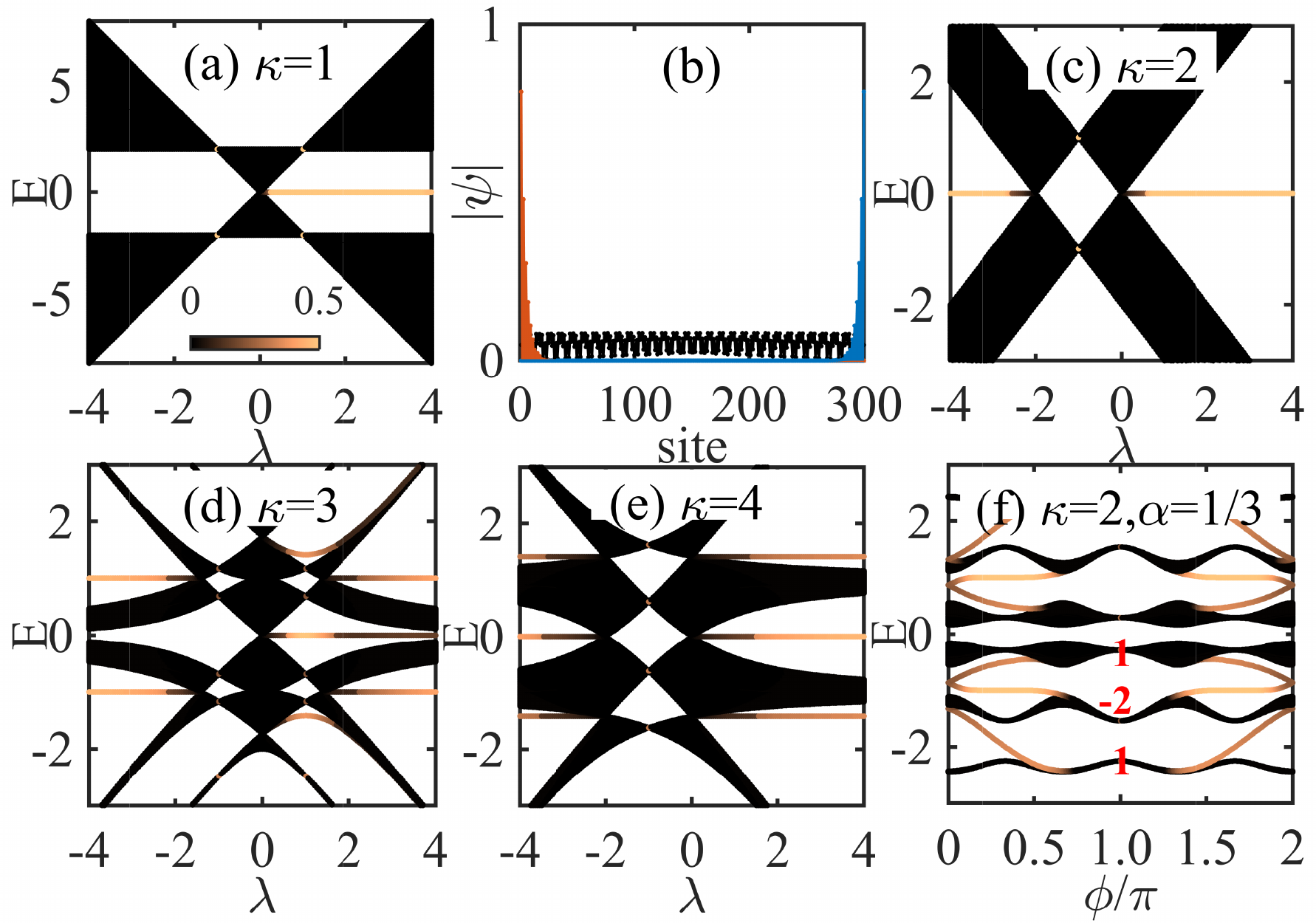}
\caption{(Color online)(a),(c)-(e) Energy spectra for the commensurate off-diagonal mosaic lattice model under open boundary conditions as a function of the mosaic modulation $\lambda$. The color bar indicates the IPR values of the eigenstates, where states with large IPR values correspond to the topological edge modes. (b) The distribution of bulk states (black) and edge states (red and blue) in the lattice. In (a)-(e) we have set $\alpha=1/2$, $\phi=0$, and $L=300$. (f) shows the energy spectrum for the system with $\kappa=2$, $\alpha=1/3$ and $\lambda=1$ as a function of $\phi$. Red numbers are the Chern numbers of the gapped bands.}
\label{fig2}
\end{figure}

If the mosaic modulation is commensurate, we can choose $\alpha=p/q$ with $p$ and $q$ being co-prime integers. When $\kappa=1$, the model reduces to the commensurate off-diagonal AAH model~\cite{Ganeshan2013PRL}, where topological zero-energy modes are observed in the gapless spectra, as shown in Fig.~\ref{fig2}(a). The color bar in the figure indicates the inverse participation ratio (IPR) value of the eigenstate, which is defined as $\text{IPR}(E_n)=\sum_j^{L} |\psi_j(E_n)|^4/[\sum_j |\psi_j(E_n)|^2]^2$. Here $\psi_j(E_n)$ represent the $j$th component of eigenstate $\psi(E_n)$ with eigenenergy $E_n$. The IPR values are of the order $O(1/L)$ for extended states but become of order $O(1)$ for localized states. In Fig.~\ref{fig2}(b), we show the distribution of eigenstates in real space. States with small IPR are bulk states spread over the whole lattice (black lines), while those with large IPR values are localized at the two ends of the 1D chain (red and blue lines). These localized states are topological edge modes. For $\kappa=2$ and $4$, all the edge modes have constant energy, while for the system with $\kappa=3$, some edge modes have non-constant energy. So by tuning the mosaic modulation, we can obtain topological edge modes in the commensurate lattices.

It should be noticed that in the case with $\alpha=1/2$ and $\kappa=3$, the width of the band gap near zero energy decreases as we increase the value of $\lambda$, as shown in Fig.~\ref{fig2}(d). However, the gap will remain finite though small and the zero edge modes always exist for $\lambda>0$. In addition, we also find that the non-constant energy edge modes in this case can exist in the continuous bulk band when $\lambda$ becomes strong enough (see details in \ref{AppendixA}).

If we generalize the model to 2D by taking $\phi$ as the momentum along another direction, we can obtain Chern insulators under appropriate parameters [see Fig.~\ref{fig2}(f)]. The gapped bands are characterized by Chern numbers defined as 
\begin{equation}
C_n=\frac{1}{2\pi}\int_{0}^{2\pi}d\phi \int_{0}^{2\pi} dk \Omega_n(k,\phi)
\end{equation}
with 
\begin{equation}
\Omega_n(k,\phi)=i(\langle \partial_k \Psi_n(k) | \partial_\phi \Psi_n(k) \rangle - \langle \partial_\phi \Psi_n(k) | \partial_k \Psi_n(k) \rangle),
\end{equation}
where $k$ is the momentum and $|\Psi_n(k) \rangle$ is the eigenstate of the Hamiltonian in momentum space. The Chern numbers for the lowest three bands are indicated by the red numbers in the figure, which are equal to the number of chiral edge states connected to the corresponding bands.

\begin{figure}[t]
\includegraphics[width=3.4in]{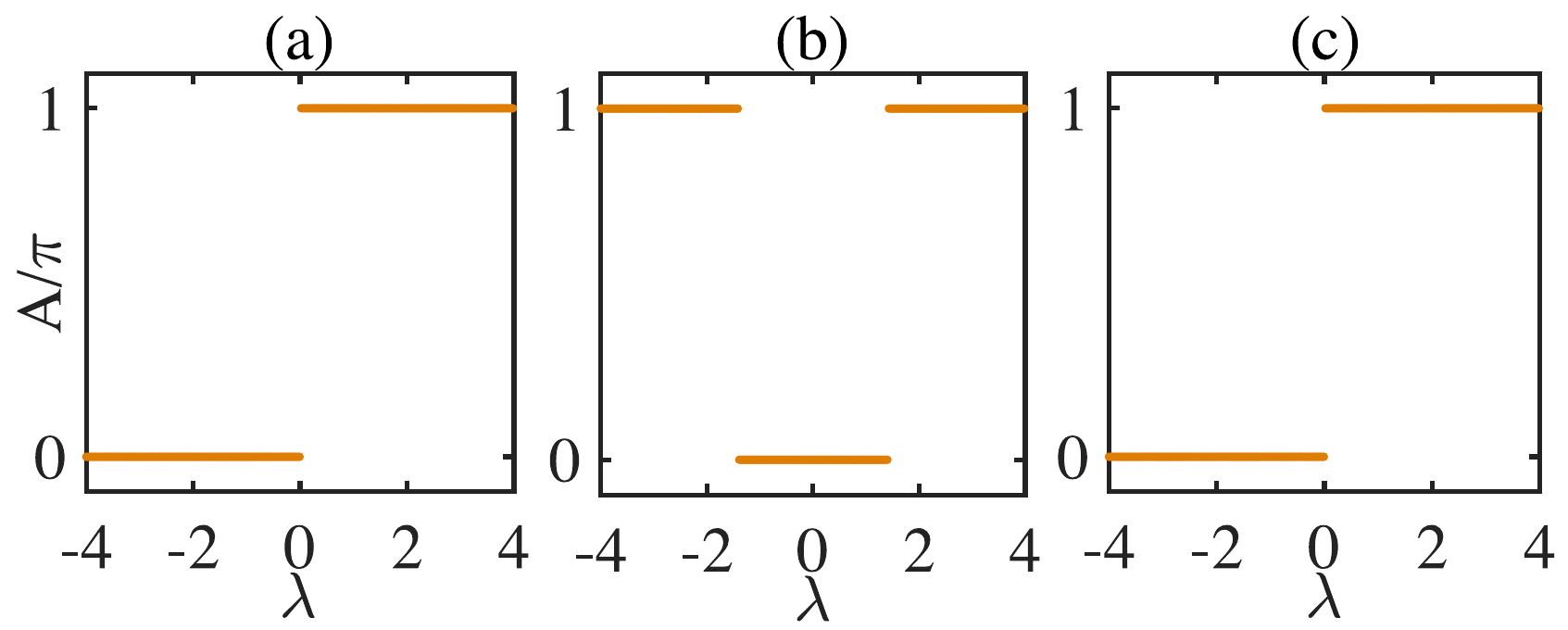}
\caption{(Color online)Berry phase as a function of $\lambda$ for the commensurate off-diagonal mosaic model with $\alpha=1/2, \kappa=3$. The Berry phase is calculated using the eigenstates in momentum space. Yellow dots represent the Berry phase for systems occupied the lowest band (a), lowest two bands (b), and lowest three bands (c).}
\label{fig3}
\end{figure} 

In all the commensurate off-diagonal mosaic models with ($q \times \kappa$) being even, zero-energy edge modes are observed due to the presence of chiral symmetry. These zero-energy edge modes are characterized by winding numbers. However, if ($q \times \kappa$) is odd, there will be no chiral symmetry in the system and the winding number cannot be well defined. In addition, the characterization of topological edge modes with nonzero energy requires a topological invariant other than winding numbers. Here we show that Berry phase can be used to characterize these nontrivial phases in a systematical way. The Berry phase for the $n$th band is defined as 
\begin{equation}
A_n = \int_{-\pi}^{\pi} dk \langle \Psi_n (k) | \partial_k \Psi_n (k) \rangle.
\end{equation}
In Fig.~\ref{fig3}, we show the numerical results of Berry phases for the lowest three bands with $E<0$ of the system with $\alpha=1/2, \kappa=3$. For convenience, we label these bands in ascending order by numbers from 1 to 3. If only the lowest band is occupied, we have $A=A_1 mod (2\pi)$. The result shown in Fig.~\ref{fig3}(a) indicates that when $\lambda>0$, the system is nontrivial, which is consistent with the regime harboring edge modes shown in Fig.~\ref{fig1}(d). For the topological edge states with energy $E=-1$, we need to add the Berry phases for the lowest two bands together and get $A=(A_1+A_2) mod (2\pi)$, as shown by the yellow dots in Fig.~\ref{fig3}(b). The Berry phase jumps from 0 to $\pi$ at around $\lambda_c = \pm 1.4$, which is exactly the critical value for the emergence of edge modes. The same method is also used to calculate the Berry phase for the zero-energy edge states, which corresponds the half-filling case. From the energy spectrum shown in Fig.~\ref{fig2}(d), we know that the zero-energy edge modes only exist for $\lambda>0$. In Fig.~\ref{fig3}(c), we can see that the Berry phase jumps sharply from $0$ to $\pi$ at $\lambda=0$, indicating a transition from trivial to nontrivial phase. Note that we have added up the Berry phases of the three occupied bands, i.e. $A=(A_1+A_2+A_3) mod (2\pi)$. If we only consider the Berry phase for the band 3, we cannot obtain the correct critical value for the phase transition. Similar rules also apply to systems with other $\kappa$ values. So, in order to get the correct topological invariant, the Berry phases of all occupied bands should be summed up together.   

The topological properties discussed above are quite similar to those appearing in the Chern insulator where the edge states are characterized by the Chern number defined over the occupied bands. The difference is that our model is 1D and  is characterized by quantized Berry phases.  

\section{Anderson localization and topological phases in incommensurate lattices}\label{sect4}
\begin{figure}[t]
\includegraphics[width=3.4in]{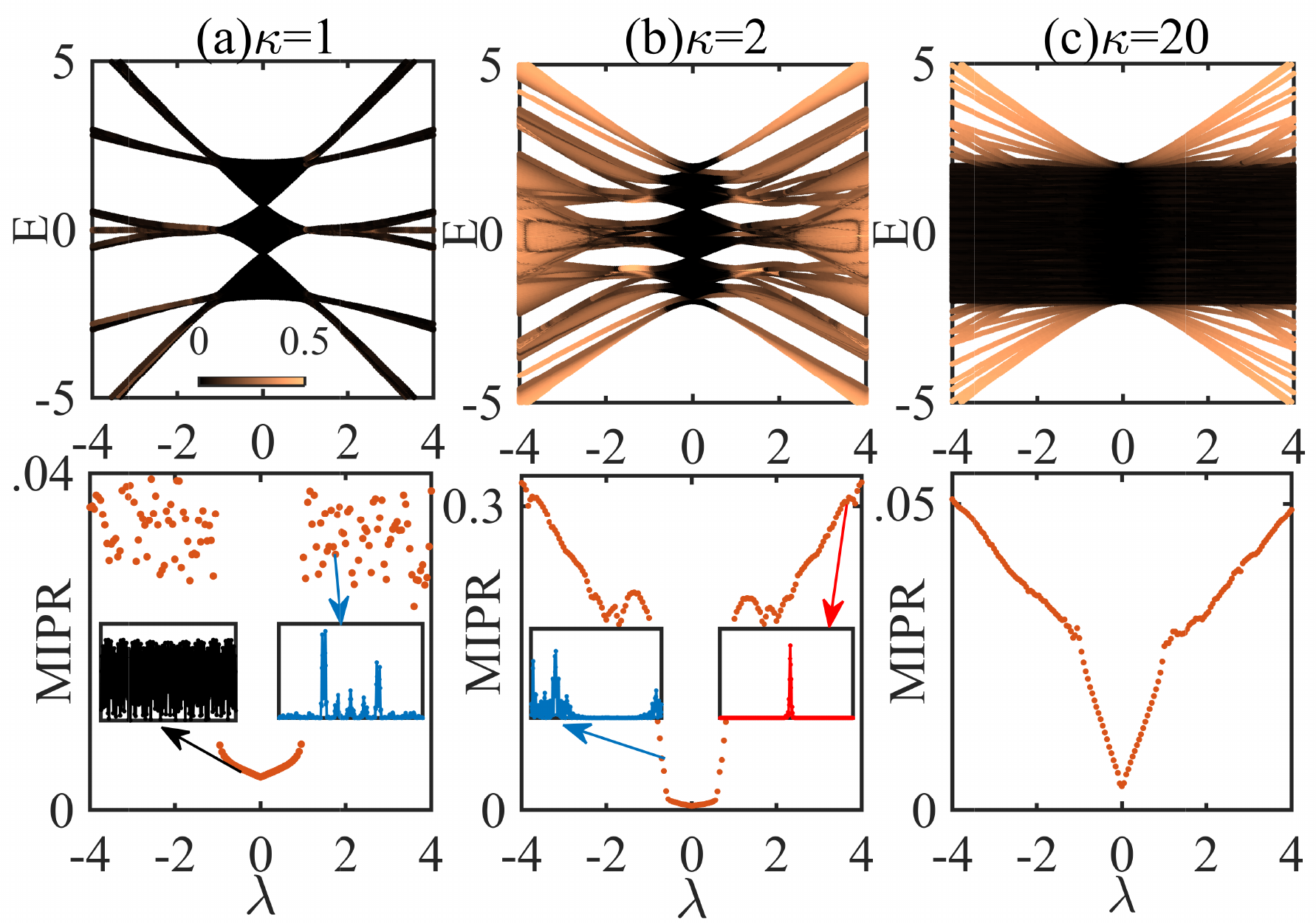}
\caption{(Color online)Upper panel: energy spectra as a function of $\lambda$ for the incommensurate off-diagonal mosaic model with different $\kappa$. Lower panel: MIPR versus $\lambda$. The insets show the distribution of eigenstates for the system with the specific $\lambda$ value indicated by the arrow. Here we use the periodic boundary condition and set $\alpha=(\sqrt{5}-1)/2$, $\phi=0$, and $L=377$.}
\label{fig4}
\end{figure}

When the mosaic modulations are incommensurate, i.e. $\alpha$ is irrational, Anderson localization will happen in the model with $\kappa > 1$. For a regular model with $\kappa=1$, the incommensurate off-diagonal modulation cannot localize the states but only make them multifractal, which can be inferred by analyzing the fractal dimensions of the eigenstates (see details in \ref{AppendixB}). Without loss of generality, we take $\alpha=(\sqrt{5}-1)/2$ in this paper. The critical value for the transition from extended phase to multifractal phase is $\lambda_c = \pm 1$. If $|\lambda| < 1$, all the states are extended, while for $|\lambda| > 1$, they becomes multifractal. We also present the value of the mean inverse participation ratio (MIPR), which is defined as $\text{MIPR}=\sum_{n=1}^L \text{IPR}(E_n)/L$. The MIPR is very close to zero and jumps to a finite but small value at $\lambda_c$ [see Fig.~\ref{fig4}(a)]. These two regimes correspond to the extended and multifractal phase, respectively. However, there is no localized state under this situation. Now if we change $\kappa$ from 1 to 2, localized states show up. From Fig.~\ref{fig4}(b), we find that MIPR is close to zero for $|\lambda|<0.55$ and the eigenstates are extended. So the regime for the extended phase shrinks comparing to the regular lattice with $\kappa=1$. When $|\lambda|>0.55$, some of the eigenstates will become localized. Notice that in order to get localized states, the off-diagonal modulation should be incommensurate as well as mosaic. If the modulation is commensurate, then the system is periodic and the eigenstates are extended. On the other hand, if the modulation is not mosaic but regular, namely, $\kappa=1$, then the states can only be multifractal.

Another variable that is very useful in characterizing the localization properties of eigenstate is the fractal dimension, which is defined as $\Gamma(E_n)=-\lim_{L\rightarrow \infty} \frac{\log[IPR(E_n)]}{\log(L)}$. In the limit $L \rightarrow \infty$, we have $\Gamma \rightarrow 1$ for extended states and $\Gamma \rightarrow 0$ for localized states. If the state is multifractal, then $0<\Gamma<1$. In ~\ref{AppendixB}, we present the numerical results of fractal dimensions for the mosaic off-diagonal model with $\kappa=1$ and $\kappa=2$ under different lattice sizes. From the different behaviors of the fractal dimension, we can conclude whether the eigenstate is extended, multifractal, or localized.  

In previous studies, the Anderson localization phenomenon has been explored in lattices with off-diagonal disorders~\cite{Economou1977SSC,Antoniou1977PRB,Weaire1977SSC,Eilmes2001PB}. By further combining with diagonal disorders, the influences of off-diagonal disorders on the localization properties have also been investigated in various models~\cite{Biswas2000PSS,Martin2011OE,Liu2017PLA}. Different from these systems, here we show that in the mosaic lattices, the Anderson localization phenomenon can be induced purely by the quasiperiodic off-diagonal modulations.   

\begin{figure}[t]
\includegraphics[width=3.4in]{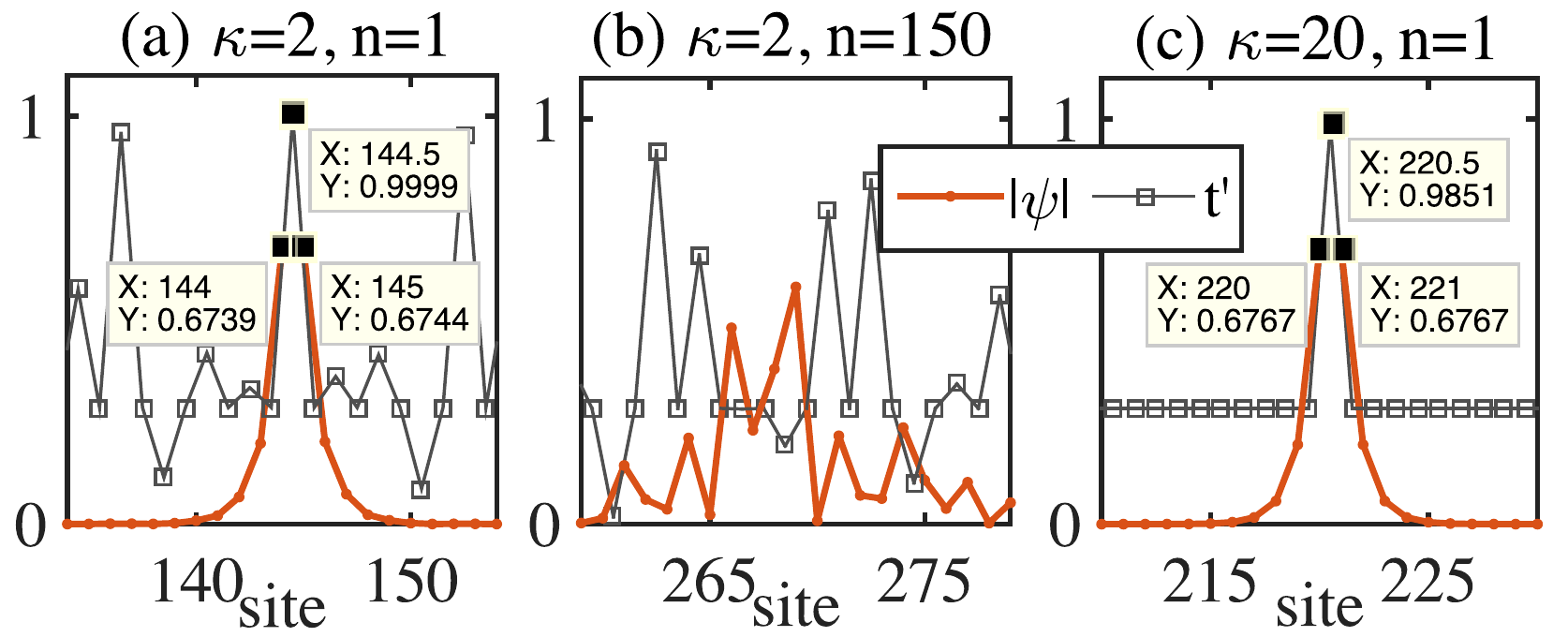}
\caption{(Color online) The distribution of localized state in the incommensurate lattice with (a)(b) $\kappa=2$ and (c) $\kappa=20$. Gray squares represent the normalized hopping amplitudes $t'$. We have set $\lambda=2.5$ and $L=377$ here. The hopping amplitudes are shifted along the horizontal axis by $0.5$. The red dots represent the distribution of the $n$th eigenstate which is sorted by the eigenenergies in ascending order.}
\label{fig5}
\end{figure}   

To get a better understanding of the localization properties, we plot the distribution of the localized states and the hopping amplitudes in Fig.~\ref{fig5}. The hopping amplitudes are normalized as  $t_{j}^{'}=|t+\lambda \cos (2\pi \alpha j+\phi)|/|t+\lambda|$ and shifted along the horizontal axis by 0.5. In Fig.~\ref{fig5}(a), the gray square at 144.5 indicates the hopping between the 144th and 145th sites in the incommensurate lattice with $\kappa=2$ and $\lambda=2.5$, which is a modulated hopping amplitude and a local maximum in the lattice. One of the eigenstates (red dots) near the band edge is localized on these two sites with almost the same distribution weight. Such localized states also exist in the system with $\kappa=20$, as shown in Fig.~\ref{fig5}(c). For the states near the band center, however, they are more extended but the main peaks still center around the sites connected by the modulated hopping terms, which are local minimums or maximums [see Fig.~\ref{fig5}(b)].  As $\lambda$ gets stronger, these states will be more localized on two neighboring sites though the distribution weights might be unequal [see \ref{AppendixC}]. The modulated hopping terms can be taken as defect links in the lattice and the sites connected by such terms play the role of impurities. An eigenstate will be stuck on them, leading to the two-site localized states. Since the mosaic modulation occurs periodically with interval $\kappa$, the number of defect links is determined by the number of quasicells which contain the nearest $\kappa$ sites. When $\kappa$ gets larger, the number of localized states will decrease since there are less impurities in the lattice, as compared in Figs.~\ref{fig4}(b) and (c). However, in a regular incommensurate lattice with $\kappa=1$, every hopping amplitude is different; the states can distribute over many sites and thus become multifractal. This simple explanation gives us a physically intuitive way to understand the Anderson localization in the incommensurate off-diagonal lattices.

\begin{figure}[t]
\includegraphics[width=3.4in]{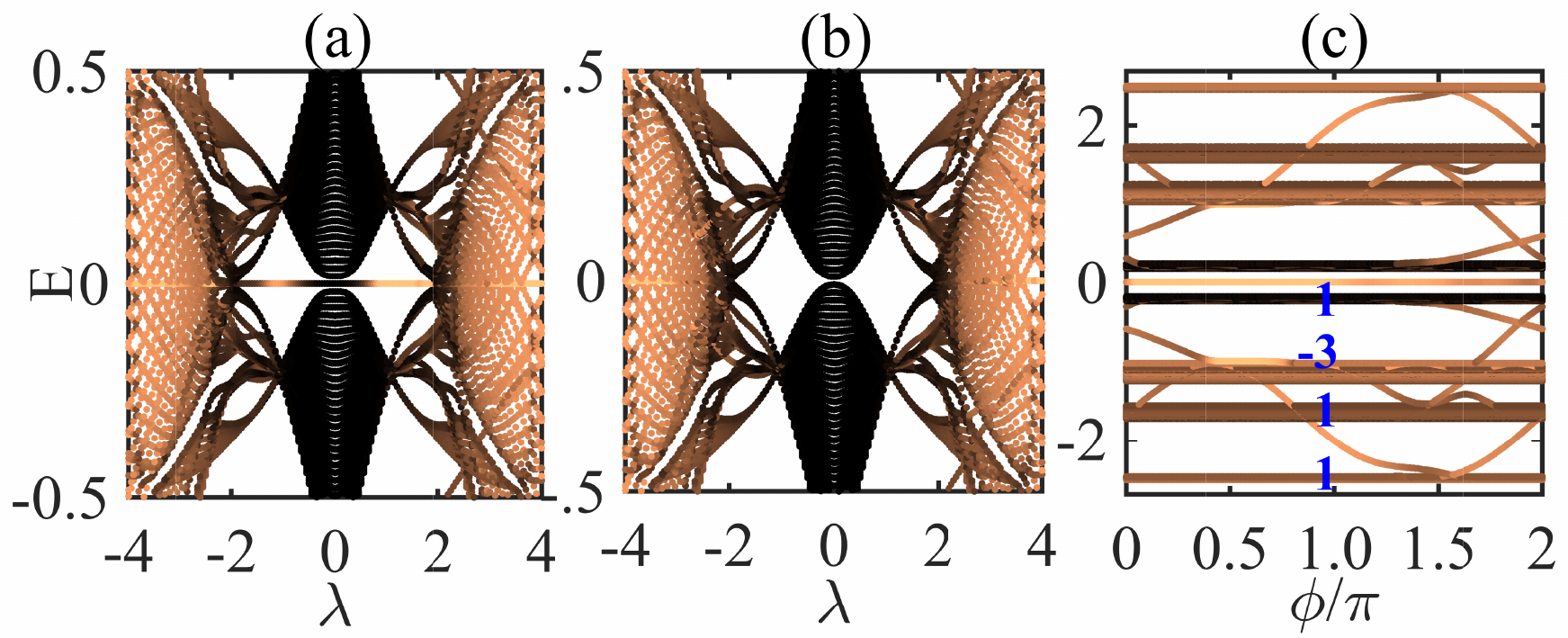}
\caption{(Color online)Energy spectra of the incommensurate lattice with $\kappa=2$ and $\alpha=(\sqrt{5}-1)/2$ under open boundary condition. (a) and (b) are the zooming in of the spectra for system with (a) $L=377$ and (b) $L=378$. Here we have chosen $\phi=0$. (c) shows the spectrum as a function of $\phi$ at $\lambda=1$, the blue number indicates the Bott index for the four bands below zero energy. The lattice size is $L=377$.}
\label{fig6}
\end{figure}

By zooming into the spectra for the lattice with $\kappa=2$ in Fig.~\ref{fig4}(b), we find there is a zero-energy mode across the whole parameter range when the lattice size is odd [see Fig.~\ref{fig6}(a)]. However, if the number of the lattices sites is even, e.g., $L=338$, the zero mode disappears [Fig.~\ref{fig6}(b)]. Such even-odd effect has been reported in the regular lattice models~\cite{Ganeshan2013PRL,Duncan2018PRB}. A similar effect also exists in the commensurate off-diagonal mosaic lattice model, where both the zero- and nonzero-energy edge modes are influenced in a more complicated way depending on whether $L$ is a multiple of $(q \times \kappa)$ or not (see \ref{AppendixD}). We further calculate the energy spectrum as a function of $\phi$, as presented in Fig.~\ref{fig6}(c). Again, gapped Chern bands exist in the 2D generalization. The Chern band is characterized by the Bott index, which is equivalent to the Chern number~\cite{LoringEPL2010,Toniolo2017arxiv} and is defined in the real space as follows
\begin{equation}
\text{Bott}=\frac{1}{2\pi} \text{Im Tr} \ln [U_y U_x U_y^\dagger U_x^\dagger].
\end{equation}
Here, $U_{a,mn}=\langle \Psi_m | e^{2\pi i \hat{a}/L_a} | \Psi_n \rangle$ with $a=x,y$, and $\hat{x}$, $\hat{y}$ denoting the position operators along the $x$ and $y$ direction, respectively. $L_a$ is the size of the system along the $a$ direction. $| \Psi_n \rangle$ represents the eigenvectors in a separable band. If we take $\phi$ as the momentum along the $y$ direction, then we can calculate the Bott index of $H_{2D}$ under periodic (open) boundary conditions along the $y$ $(x)$ direction. The Bott indices for the gapped bands are shown by the blue numbers in the figure. So, in the incommensurate off-diagonal mosaic lattices, we can also obtain topological Chern insulators.

\section{Summary}\label{sect5}
In summary, we have investigated the topological phases and Anderson localization in the one-dimensional off-diagonal mosaic lattices. When the lattice is commensurate, we find that topological edge modes with zero and nonzero energies can exist by tuning the modulations. The nontrivial regimes are characterized by quantized Berry phases which are calculated by summing up the phases for all the occupied energy bands. If the lattice becomes incommensurate, the Anderson localization phenomenon will show up. The eigenstate is found to be centered on two neighboring sites connected by the mosaic modulated hopping terms. In addition, we also find that our model can host Chern insulators when generalized to 2D, both in the commensurate and incommensurate cases.

In Ref.~\cite{Kraus2012PRL}, Kraus \emph{et al.} demonstrate that the 1D off-diagonal AAH model can be realized by using coupled optical waveguides. Such method can be readily adopted to realize the off-diagonal mosaic lattice model, and all the phenomena reported in this paper could be examined. In addition, as a new and versatile platform for simulating topological systems, electrical circuits have been extensively used to realize various lattice models recently~\cite{Lee2018CommPhys,Hofmann2019PRL}. Our model is also reliable by using LC circuits, where the system's properties are detectable from the impedance measurements.  

\section*{Acknowledgments}
R. L. is supported by NSFC under Grant No. 11874234 and the National Key Research and Development Program of China (2018YFA0306504).

\section*{Appendix}\label{Appendix}
\setcounter{equation}{0} 
\renewcommand{\theequation}{{A}\arabic{equation}}
\setcounter{section}{0} 
\renewcommand{\thesection}{{APPENDIX }\Alph{section}}

In this Appendix, we will first discuss the details of the energy spectra and the existence of topological edge modes for the 1D off-diagonal mosaic lattices with large $\lambda$ values. Then we present the numerical results of the fractal dimensions for the incommensurate lattices with different sizes. In addition, the distributions of localized states and hopping amplitudes in the incommensurate lattice are analyzed. Finally, we present the energy spectra for the systems with different lattice sizes, where the changing of the size results in the even-odd effect in the edge modes of the system. 

\section{Energy gap and edge modes in systems with strong $\lambda$}\label{AppendixA}
\begin{figure}[h]
  \includegraphics[width=3.4in]{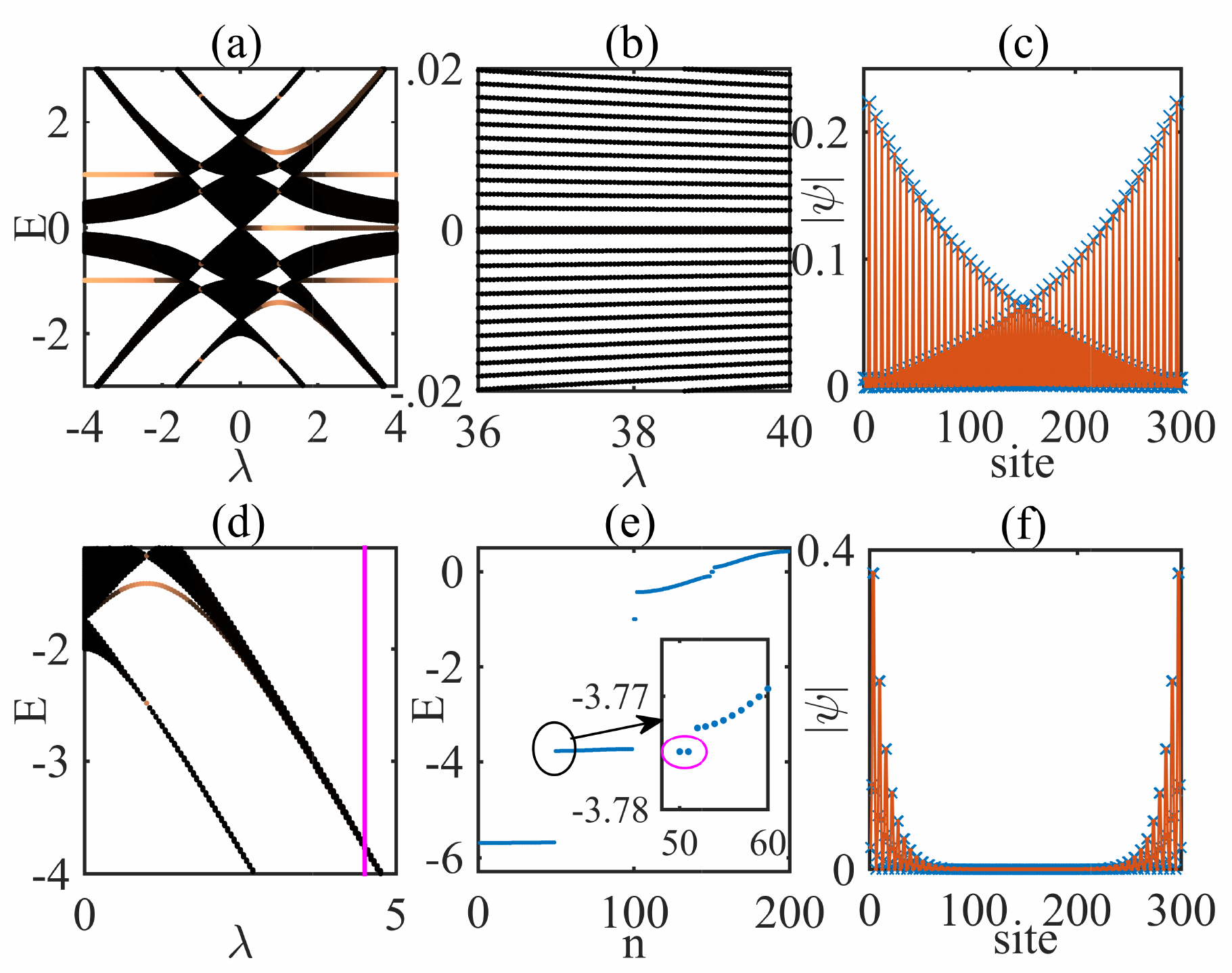}
  \caption{(Color online) Energy spectrum and edge states for the off-diagonal mosaic lattice with $\alpha=1/2$ and $\kappa=3$. (a) shows the spectrum as a function of $\lambda$ with $\lambda \in [0 4]$. (b) is the enlargement of the spectrum near zero energy for stronger $\lambda$ values. The zero-energy modes in the gap are shown in (c). (d) is an enlargement of the spectrum near the lowest two bands. The energy levels at $\lambda=4.5$ indicated by the magenta line is shown in (e), where the edge modes in the magenta circle correspond to the two edge modes shown in (f).}
\label{fig7}
\end{figure}

In the main text, we have shown that for the system with $\alpha=1/2$ and $\kappa=3$, there are zero- as well as nonzero-energy edge modes. For the zero-energy modes, the gap near the zero energy decreases as $\lambda$ increases, but stays finite even when $\lambda$ becomes very strong, as shown in Fig.~\ref{fig7}(a). It seems that when $\lambda$ is strong enough, the gap will close and zero modes will disappear. However, we have calculated the spectrum for system with $\lambda$ up to $40$, where the gap is found to be finite though very small [see Fig.~\ref{fig7}(b)]. And we can still observe two zero modes localized at the two ends of the system [Fig.~\ref{fig7}(c)]. So, the band gap near zero energy will remain finite and zero-energy edge modes will always exist in this case, which is consistent with the results obtained from the Berry phase, which shows that the system is nontrivial for $\lambda>0$ at half-filling (see Fig.~\ref{fig3} in the main text).

Next we turn to the nonzero-energy edge modes near the lowest energy band. As can be seen from Fig.~\ref{fig7}(d), the energy of the edge modes is not constant. As $\lambda$ grows, the edge modes will get closer and closer to the continuous bulk band, as shown in Fig.~\ref{fig7}(e). Interestingly, the edge modes still exist in the continuum of extended bulk states and thus the system is nontrivial for $\lambda>0$ for the lowest energy band. This is again consistent with the nontrivial regime revealed from the Berry phase discussed in the main text.

\section{Fractal dimensions of localized, extended and multifractal states}\label{AppendixB}
To characterize the localized property of the eigenstates, we can also use the fractal dimension, which is defined as
\begin{equation}
\Gamma(E_n) = - \lim_{L \rightarrow \infty} \frac{\log[IPR(E_n)]}{\log(L)}.
\end{equation}
Here $IPR(E_n)$ is the inverse participation ratio of the eigenstate with eigenenergy $E_n$. For localized states, we have $\Gamma \rightarrow 0$, while for extended states, we have $\Gamma \rightarrow 1$. If the states are multifractal, then the corresponding fractal dimension will be a finite number between $0$ and $1$, i.e., $0<\Gamma<1$.

\begin{figure}[h]
  \includegraphics[width=3.4in]{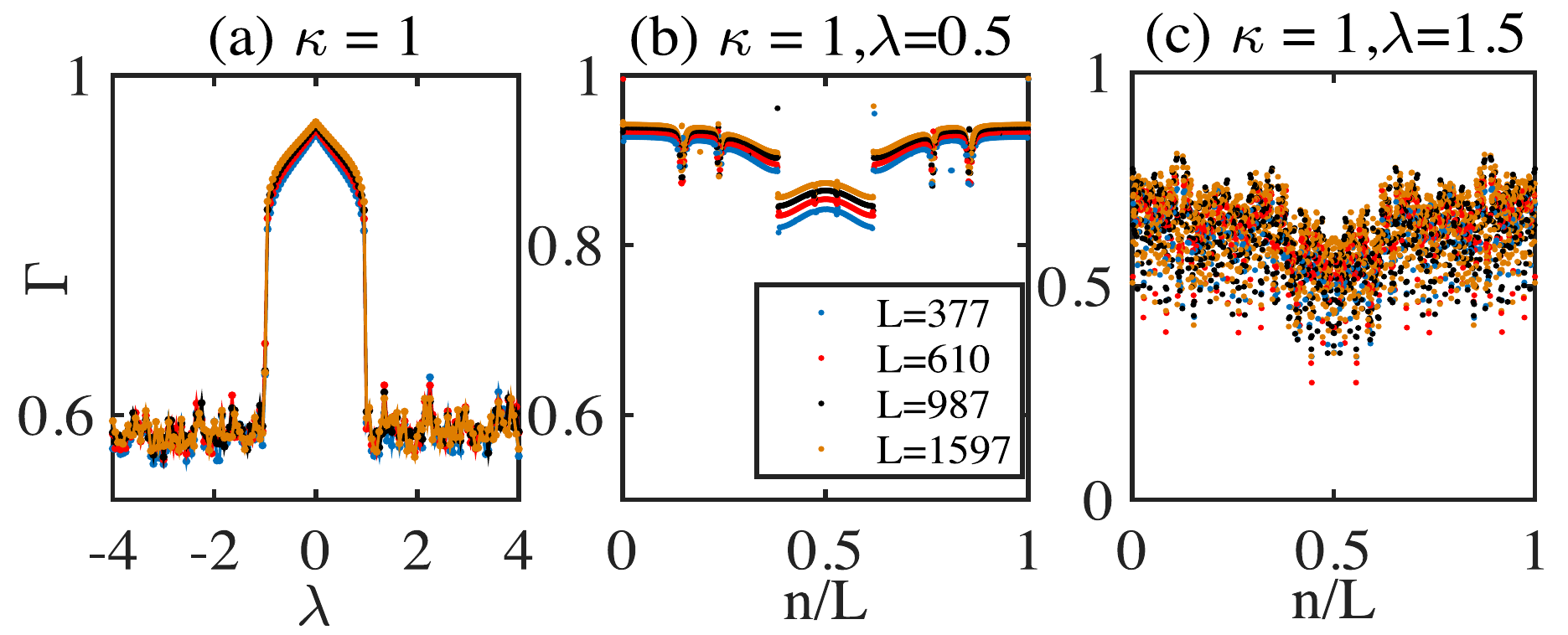}
  \caption{(Color online) Fractal dimensions for the states of 1D mosaic off-diagonal lattice with $\kappa=1$. (a) shows the fractal dimension as a function of $\lambda$, which is obtained by averaging over all the eigenstates in the system. (b) and (c) show the fractal dimension of each eigenstate at $\lambda=0.5$ and $1.5$, respectively. The different dots represent the systems with different lattice sizes.}
\label{fig8}
\end{figure}

In Fig.~\ref{fig8}, we present the fractal dimensions for the mosaic off-diagonal lattice with $\kappa=1$. The averaging fractal dimension for the system is shown in Fig.~\ref{fig8}(a), which is calculated as $\Gamma=\sum_n \Gamma(E_n)/L$. We can see that $\Gamma$ drops sharply at $|\lambda|=1$ from around $0.9$ to around $0.55$. In Figs.~\ref{fig8}(b) and (c), we present the fractal dimension of each eigenstate for the system with $\lambda=0.5$ and $\lambda=1.5$, respectively. As the lattice size increases, the fractal dimension will increase toward $1$ for the states in $|\lambda|<1$, while for the states in $|\lambda|>1$, the fractal dimension will fluctuate around $0.55$. These results indicate that the states are extended in the regime with $|\lambda|<1$ and multifractal in the regime with $|\lambda|>1$.

\begin{figure}[h]
  \includegraphics[width=3.4in]{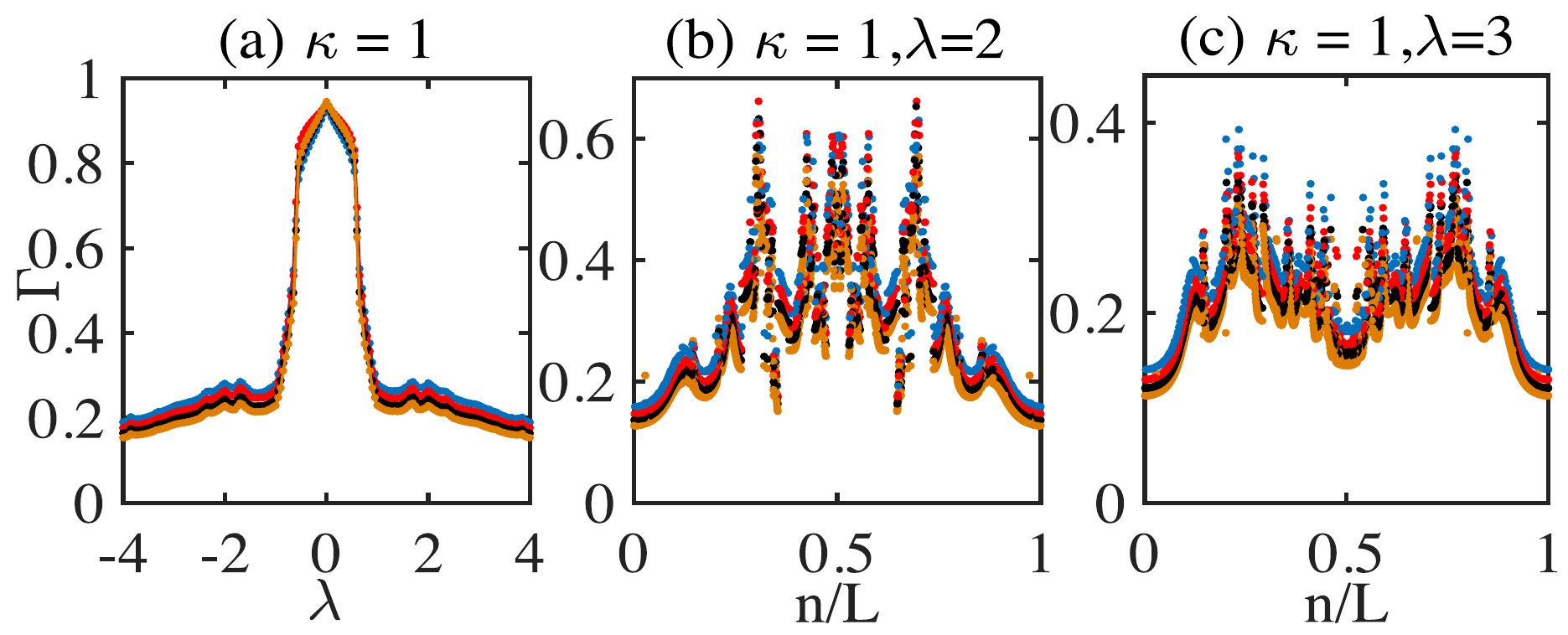}
  \caption{(Color online) Fractal dimensions for the states of 1D mosaic off-diagonal lattice with $\kappa=2$. (a) shows the fractal dimension as a function of $\lambda$. (b) and (c) show the fractal dimension of each eigenstate at $\lambda=2$ and $3$, respectively. The different dots represent the systems with different lattice size shown in Fig.~\ref{fig8}.}
\label{fig9}
\end{figure} 

Next, we discuss the behavior of the fractal dimension for the system with $\kappa=2$. The averaging fractal dimension is calculated as a function of $\lambda$, as shown in Fig.~\ref{fig9}(a). When $\lambda$ is small, the fractal dimension is close to $1$, indicating that most of the eigenstates are extended. As $\lambda$ grows, the fractal dimension will decreases gradually and drop to a value smaller than $0.2$, which implies that the eigenstates become localized. In Figs.~\ref{fig9} (b) and (c), we show the fractal dimension for each eigenstate of the system at $\lambda=2$ and $3$, respectively. We can see that the fractal dimensions for the states near the band edges are very small, while those for the states near the band center fluctuates. Furthermore, as the lattice size increases, the fractal dimension for the localized states will further decrease. Thus we conclude that when $\lambda$ is strong, the states near the band edges will be localized. Near the band center, the fractal dimensions fluctuate in a wide range ($0.2~0.65$), which implies that there are different kinds of eigenstates. 

\section{Localized states in the incommensurate off-diagonal mosaic lattice}\label{AppendixC}
\begin{figure}[h]
  \includegraphics[width=3.4in]{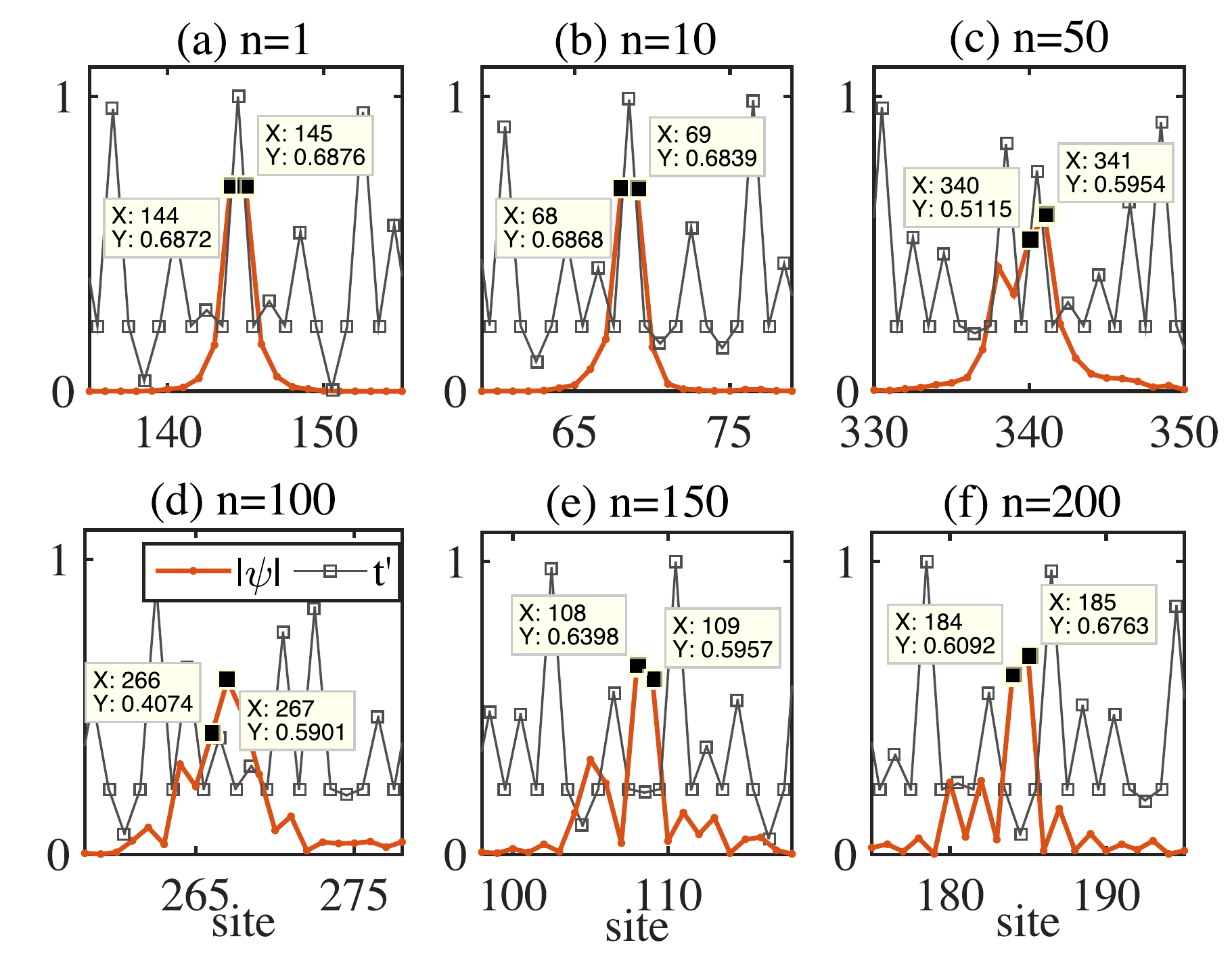}
  \caption{(Color online) The distribution of eigenstates (red dots) and hopping amplitude (gray squares) in the real space for the incommensurate off-diagonal mosaic lattice with $\alpha=(\sqrt{5}-1)/2$ and $\kappa=2$. $n$ indicates the $n$th eigenstate of the system which are sorted by eigenenergies in ascending order. Here we have set $\lambda=3.5$ and $L=377$.}
\label{fig10}
\end{figure}

Here we plot the distributions of eigenstates for the incommensurate off-diagonal mosaic lattices with $\kappa=2$ and $\lambda=3.5$ (see Fig.~\ref{fig10}). For the lattice with $L=377$, there are 377 eigenstates. We show the distribution of the normalized hopping amplitude and six different eigenstates. For the states near the band edge, the distribution weights are almost the same on two neighboring sites, as shown in (a) and (b) for the $n=1$ and $10$ eigenstate, respectively. These sites are connected by the modulated hopping terms which are local minimum or maximum in the lattice. As we move away from the band edge to the band center, the distribution weight on the two sites will become different, as can be seen in Figs.~\ref{fig10}(c)-\ref{fig10}(f). Though the states become more extended, the main peak of the state still centers around the two lattice sites which are connected by the modulated hopping terms.

\section{Even-odd effect in the edge modes}\label{AppendixD}
\begin{figure}[h]
  \includegraphics[width=3.4in]{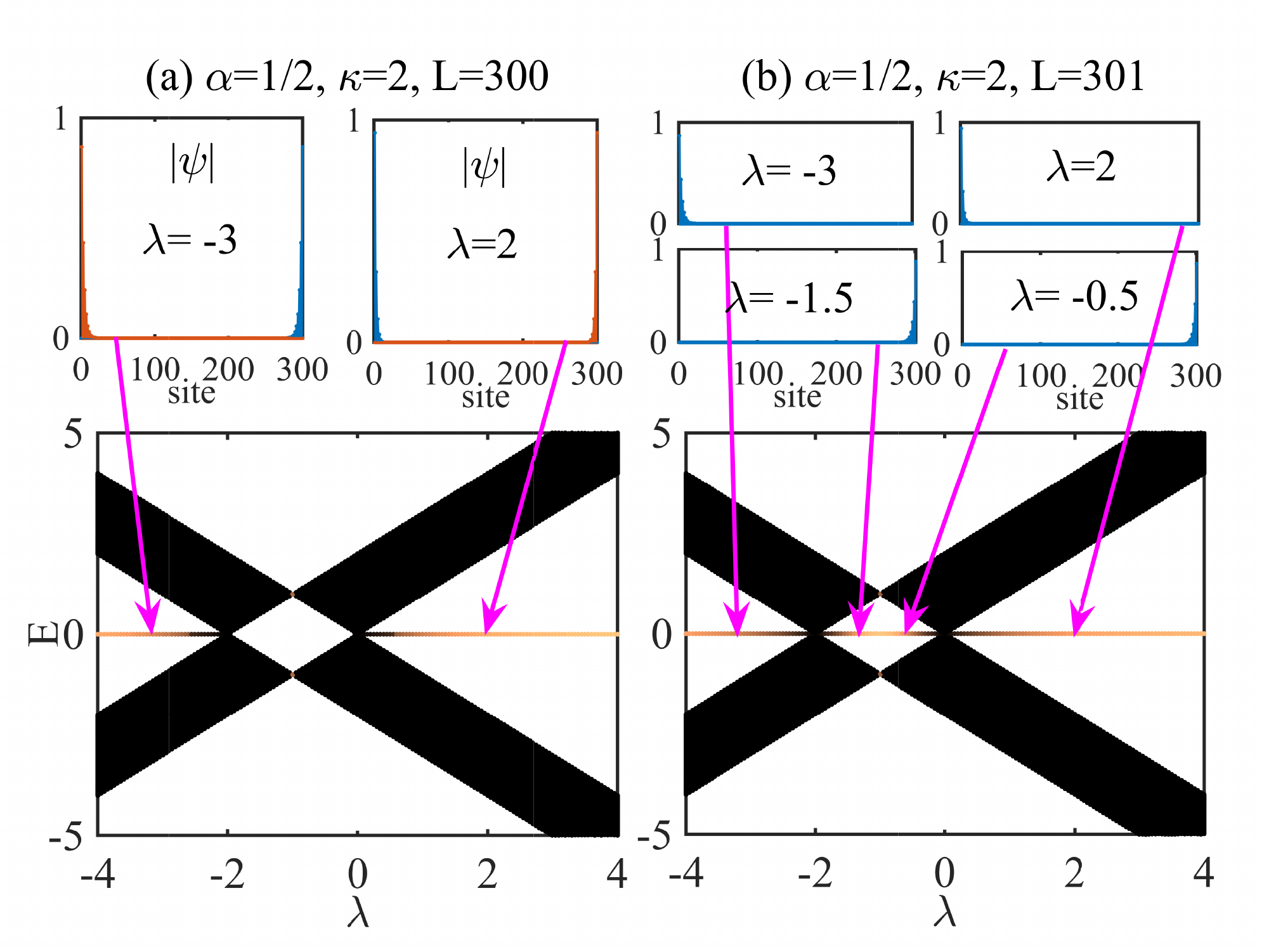}
  \caption{(Color online)Even-odd effect in the off-diagonal mosaic lattices with $\alpha=1/2$ and $\kappa=2$. (a) shows the spectrum under open boundary conditions when $L=300$ while (b) is the spectrum for the system with $L=301$. The color indicates the inverse participation ratio (IPR) of the eigenstate. The insets show the distributions of the  edge modes at specific $\lambda$ values indicated by the arrows.}
\label{fig11}
\end{figure}

In Fig.~\ref{fig11}, we show the spectra of the system with $\alpha=1/2,\kappa=2$ with different lattice sizes. When $L$ is even, the topological zero modes will show up in the nontrivial regime with $\lambda>0$. There are always two zero-energy modes in this case, which are localized at the two ends of the 1D lattice, as shown in Fig.~\ref{fig11}(a). However, if the lattice size becomes odd, then the zero mode presents in both the trivial and nontrivial regimes. In addition, there is only one single zero mode instead two as in the case with even $L$ [see Fig.~\ref{fig11}(b)]. The zero mode is localized at the left end of the lattice for $\lambda<-2$ and $\lambda>0$, but becomes localized at the right end for $-2<\lambda<0$. Such even-odd effect has also been reported in the off-diagonal AAH model~\cite{Ganeshan2013PRL}. 

\begin{figure}[h]
  \includegraphics[width=3.0in]{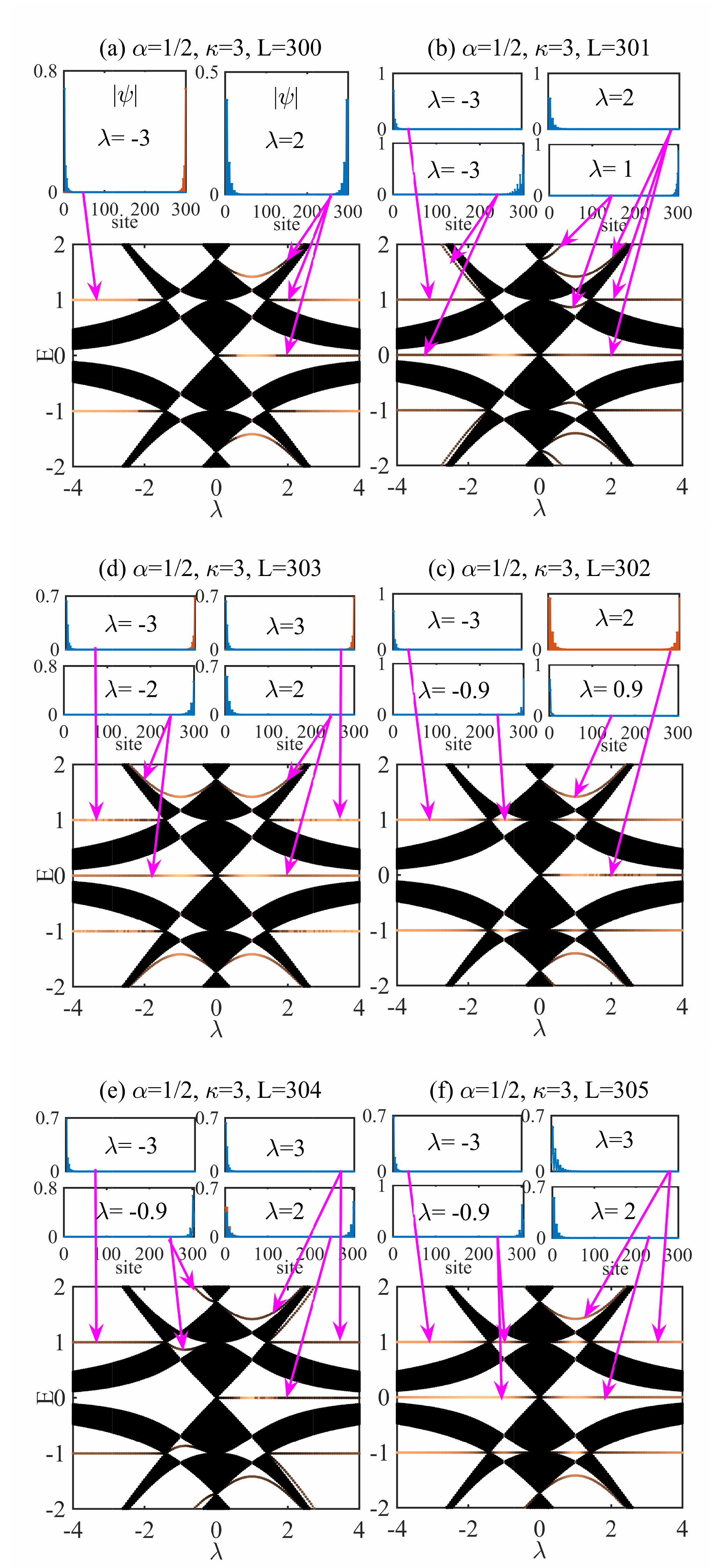}
  \caption{(Color online) Energy spectra and edge modes for the off-diagonal mosaic lattices with $\alpha=1/2$ and $\kappa=3$. (a) The edge modes in the nontrivial regimes come in pairs when $L$ is a multiple of $q \times \kappa$. (b)-(f) While if $L$ is not a multiple of $(q \times \kappa)$, the edge mode will also appear in the trivial regime, and only one single edge mode is found to be localized at one end of the lattice, as shown in (b)-(f).}
\label{fig12}
\end{figure}

The situation becomes much more complicated for the system with larger $\kappa$.For the commensurate lattice with $\alpha=p/q$, where $p$ and $q$ are coprime integers, we find that when $L$ is a multiple of $(q \times \kappa)$, all the topological edge modes come in pairs in the nontrivial regime. Figure~\ref{fig12}(a) presents the energy spectra for the system with $\alpha=1/2$, $\kappa=3$, and $L=300$, where two edge modes are always observed at every specific $\lambda$ value, which are localized at the two ends of the 1D lattice. However, if $L$ is not a multiple of $(q \times \kappa)$, the edge mode will exist in the trivial regime. Moreover, there is only one single mode localized at one end of the lattice, as can be seen in Figs.~\ref{fig12}(b)-\ref{fig12}(f). Same phenomena can also be observed in systems with other values of $\alpha$ and $\kappa$. So, in the off-diagonal mosaic lattice, richer phenomena exist due to the presence of mosaic modulations. 

\end{document}